\documentclass[12pt]{article}
\usepackage{epsf}
\usepackage[utf8]{inputenc}
\usepackage[english]{babel}

\begin{document}

\title{
{\bf Two-Pomeron eikonal approximation for the high-energy elastic diffractive scattering of nucleons}}
\author{A.A. Godizov\thanks{E-mail: anton.godizov@gmail.com}\\
{\small {\it A.A. Logunov Institute for High Energy Physics}},\\ {\small {\it NRC ``Kurchatov Institute'', 142281 Protvino, Russia}}}
\date{}
\maketitle

\vskip-1.0cm

\begin{abstract}
It is demonstrated that the elastic diffractive scattering of nucleons at collision energies higher than 540 GeV and transferred momenta lower than 2 GeV, including the 
Coulomb-nuclear interference region, can be described in the framework of a very simple Regge-eikonal model where the eikonal is just a sum of two supercritical Regge pole 
terms. The predictive value of the proposed approximation is verified.
\end{abstract}

\section*{1. Introduction}

During the last several decades, perturbative quantum chromodynamics (pQCD) confirmed its usefulness many times as a powerful theoretical tool in the sector of high energies 
and high transferred momenta of strongly interacting particles. However, at present, a very large part of hadron physics cannot be treated in the framework of this 
quantum-field model. Particularly, to describe quantitatively the elastic diffractive scattering (EDS) of nucleons at high values of the collision energy and low values of 
the transferred momentum, we have to use phenomenological models which are not based on any analytic approximations within QCD. The absence of a direct connection between these 
models and the fundamental theory of strong interaction very often reduces their predictive power. As a result, in 2011, many hadron diffraction models nicely described the 
available experimental data on the proton-(anti)proton EDS in the energy range from the ISR to Tevatron (with the collision energy increase in several tens of times), but 
failed completely to reproduce the behavior of the $pp$ angular distribution in the region of the diffraction dip and nonforward peak at the LHC \cite{totem7}. The 
discrepancy between the model predictions and the experimental data was so huge that it could lead to a justified conclusion about our total misunderstanding of the physical 
mechanisms of hadron diffraction (for detailed discussion, see a minireview in \cite{godizov0}).

Nonetheless, the fraction of EDS events in the total number of events at high-energy hadron colliders is so significant ($\sim$ 25\% at the LHC) that we still strongly need 
to develop reliable phenomenological approaches which could help to intepret properly the results of the measurements in the high-energy hadron diffraction region. Of course, 
any model should be verifiable (and, certainly, discriminable) on the available and forthcoming experimental data.

All the modern models of EDS could be divided into two groups: those ones based on Regge theory \cite{collins} and the non-Reggeon models. The references to various Reggeon 
and non-Reggeon phenomenological schemes can be found in a recent review \cite{jenkovszky} and in the RPP \cite{rpp}. The aim of this paper is to demonstrate the predictive 
value and reliability of the two-Pomeron eikonal approximation proposed earlier in \cite{godizov}. It will be done via application to the new elastic scattering data sets 
produced recently by the TOTEM Collaboration \cite{totem13}.

\section*{2. The model}

\subsection*{2.1. The strong interaction subamplitude}

The physical content of the two-Pomeron eikonal approximation is very simple and transparent. In the kinematic range wherein the Coulomb interaction may be considered 
negligible, the eikonal for the high-energy EDS of nucleons can be represented as a sum of two supercritical Regge pole terms. The first term corresponds to exchange by the 
so-called soft Pomeron (SP). This interaction is the basic cause of the visible growth of the proton-proton total and elastic cross-sections at high energies. The second term 
corresponds to exchange by the hard Pomeron (HP), also known as the BFKL Pomeron. A detailed discussion of the considered model can be found in \cite{godizov}. Below we just 
give a recipe for calculation of the EDS angular distributions in the framework of this approximation:
$$
\frac{d\sigma}{dt} = \frac{|T_N(s,t)|^2}{16\pi s^2}\,\;,\;\;\;\;T_N(s,t) = 4\pi s\int_0^{\infty}d(b^2)\,J_0(b\sqrt{-t})\,\frac{e^{2i\delta_N(s,b)}-1}{2i}\;,
$$
\begin{equation}
\label{eikrepr}
\delta_N(s,b)=\frac{1}{16\pi s}\int_0^{\infty}d(-t)\,J_0(b\sqrt{-t})\,[\delta_{\rm SP}(s,t) + 
\delta_{\rm HP}(s,t)] = \frac{1}{16\pi s}\int_0^{\infty}d(-t)\,J_0(b\sqrt{-t})\,\times
\end{equation}
$$
\times\,\left[\xi_+(\alpha_{\rm SP}(t))\,g^2_{\rm SP}(t)\,\pi\alpha'_{\rm SP}(t)\left(\frac{s}{2s_0}\right)^{\alpha_{\rm SP}(t)} + 
\xi_+(\alpha_{\rm HP}(0))\,\beta_{\rm HP}(t)\left(\frac{s}{2s_0}\right)^{\alpha_{\rm HP}(0)}\right],
$$
where $s$ and $t$ are the Mandelstam variables, $b$ is the impact parameter, $s_0 = 1$ GeV$^2$, $J_0(x)$ is the Bessel function, $\alpha_{\rm SP}(t)$ is the SP Regge 
trajectory, $g_{\rm SP}(t)$ is the SP coupling to nucleon, $\alpha_{\rm HP}(0)$ is the intercept of the HP Regge trajectory (it was argued in \cite{godizov} why the 
$t$-dependence of $\alpha_{\rm HP}(t)$ is negligible in the region of EDS), $\beta_{\rm HP}(t)\equiv g^2_{\rm HP}(t)\,\pi\alpha'_{\rm HP}(t)$ is the HP Regge residue, 
$\xi_+(\alpha)=\left(i+{\rm tan}\frac{\pi(\alpha-1)}{2}\right)$ are the signature factors for even Reggeons, $\delta_N$ is the eikonal (Born amplitude), and $T_N$ is the full 
amplitude related to strong interaction. 

\begin{table}[ht]
\begin{center}
\begin{tabular}{|l|l|}
\hline
\bf Parameter          & \bf Value         \\
\hline
$\alpha_{\rm SP}(0)-1$  & 0.109            \\
$\tau_a$                & 0.535 GeV$^2$    \\
$g_{\rm SP}(0)$         & 13.8 GeV         \\
$a_g$                   & 0.23 GeV$^{-2}$  \\
$\beta_{\rm HP}(0)$     & 0.08             \\
$b$                     & 1.5  GeV$^{-2}$  \\
$\alpha_{\rm HP}(0)-1$  & 0.32 (FIXED)     \\
\hline
\end{tabular}
\end{center}
\vskip -0.2cm
\caption{The parameter values for (\ref{pomeron}) obtained earlier \cite{godizov2,godizov} via fitting to the elastic scattering data in the collision energy range 
546 GeV $\le\sqrt{s}\le$ 7 TeV.}
\label{tab1}
\end{table}
The HP intercept can be extracted from the data on the proton unpolarized structure function $F^p_2(x,Q^2)$ \cite{struc} at high values of the incoming photon virtuality 
$Q^2$ and low values of the Bjorken scaling variable $x$: $\alpha_{\rm HP}(0)\approx 1.32$ \cite{godizov4}. In the region of low negative $t$, the unknown functions 
$\alpha_{\rm SP}(t)$, $g_{\rm SP}(t)$, and $\beta_{\rm HP}(t)$ may be approximated with the help of the following simple test parametrizations 
\cite{godizov2,godizov}: 
\begin{equation}
\label{pomeron}
\alpha_{\rm SP}(t) = 1+\frac{\alpha_{\rm SP}(0)-1}{1-\frac{t}{\tau_a}}\;,\;\;\;\;g_{\rm SP}(t)=\frac{g_{\rm SP}(0)}{(1-a_gt)^2}\;,\;\;\;\;
\beta_{\rm HP}(t)=\beta_{\rm HP}(0)\,e^{b\,t}\,,
\end{equation}
where the free parameters take on the values presented in Table \ref{tab1}.

Usage of essentially nonlinear parametrization for the SP Regge trajectory distinguishes the proposed model from other models \cite{landshoff}--\cite{martynov} which exploit 
linear Regge trajectories. The main reason to presume a strongly nonlinear behavior of $\alpha_{\rm SP}(t)$ at $t<0$ is based 
on the assumption that in the region of extremely high transfers, $\sqrt{-t}\gg 1$ TeV, any exchange by the SP turns into the exchange by two gluons. At asymptotically 
high energies, such perturbative amplitudes behave as $T^{(gg)}(s,t)\sim$ $T^{(\gamma\gamma)}(s,t)\sim s$ \cite{wu,kearney,kirschner}, what implies the following asymptotic 
relation: $\lim_{t\to -\infty}\alpha_{\rm SP}(t) = 1$. A detailed discussion of why any hadron Regge trajectory must exhibit a nonlinear behavior at $t<0$ and 
how such a nonlinearity may be consistent with approximately linear behavior in the resonance region, $t>0$, can be found in \cite{petrov2}. Unfortunately, at its current 
stage of development, QCD does not allow direct calculation of $\alpha_{\rm SP}(t)$, as well as of $g_{\rm SP}(t)$ and $\beta_{\rm HP}(t)$, in the nonperturbative region. As 
a consequence, one has to use some test parametrizations to approximate the unknown functions $\alpha_{\rm SP}(t)$, $g_{\rm SP}(t)$, and $\beta_{\rm HP}(t)$ in the range of 
low negative $t$. In view of this fact, the author must emphasize that expressions (\ref{pomeron}) are just phenomenological. In 
particular, the true Regge trajectories of both the Pomerons have branching points at the two-pion threshold (for details, see chapter 3 in \cite{collins}) and have no poles 
on the physical sheet. Therefore, parametrizations (\ref{pomeron}) must be treated as phenomenological nonanalytic approximations to the true functions 
$\alpha_{\rm SP}(t)$, $g_{\rm SP}(t)$, and $\beta_{\rm HP}(t)$ in the region $t<0$ only. They are absolutely invalid in the region ${\rm Re}\,t>0$. However, exploitation of 
such parametrizations is quite justified due to the above-mentioned fact that the exact analytic behavior of $\alpha_{\rm SP}(t)$, $g_{\rm SP}(t)$, and $\beta_{\rm HP}(t)$ 
is still unknown.

\subsection*{2.2. The impact of electromagnetic interaction}

To describe the EDS of protons in the region of Coulomb-nuclear interference we need to take account of electromagnetic interaction.

In the framework of the eikonal approach, the full amplitude of the proton-(anti)proton EDS in the coordinate representation has the following structure:
\begin{equation}
\label{elnuc}
T(s,b) = \frac{e^{2i(\delta_C(s,b)+\delta_N(s,b))}-1}{2i}=T_N(s,b)+\delta_C(s,b)+2i\,T_N(s,b)\,\delta_C(s,b) + O(\alpha_e^2)\,,
\end{equation}
where $\delta_C(s,b)\sim \alpha_e$ is the tree-level subamplitude of electromagnetic interaction, and $\alpha_e$ is the fine structure constant. 
\begin{figure}[ht]
\epsfxsize=8.2cm\epsfysize=8.2cm\epsffile{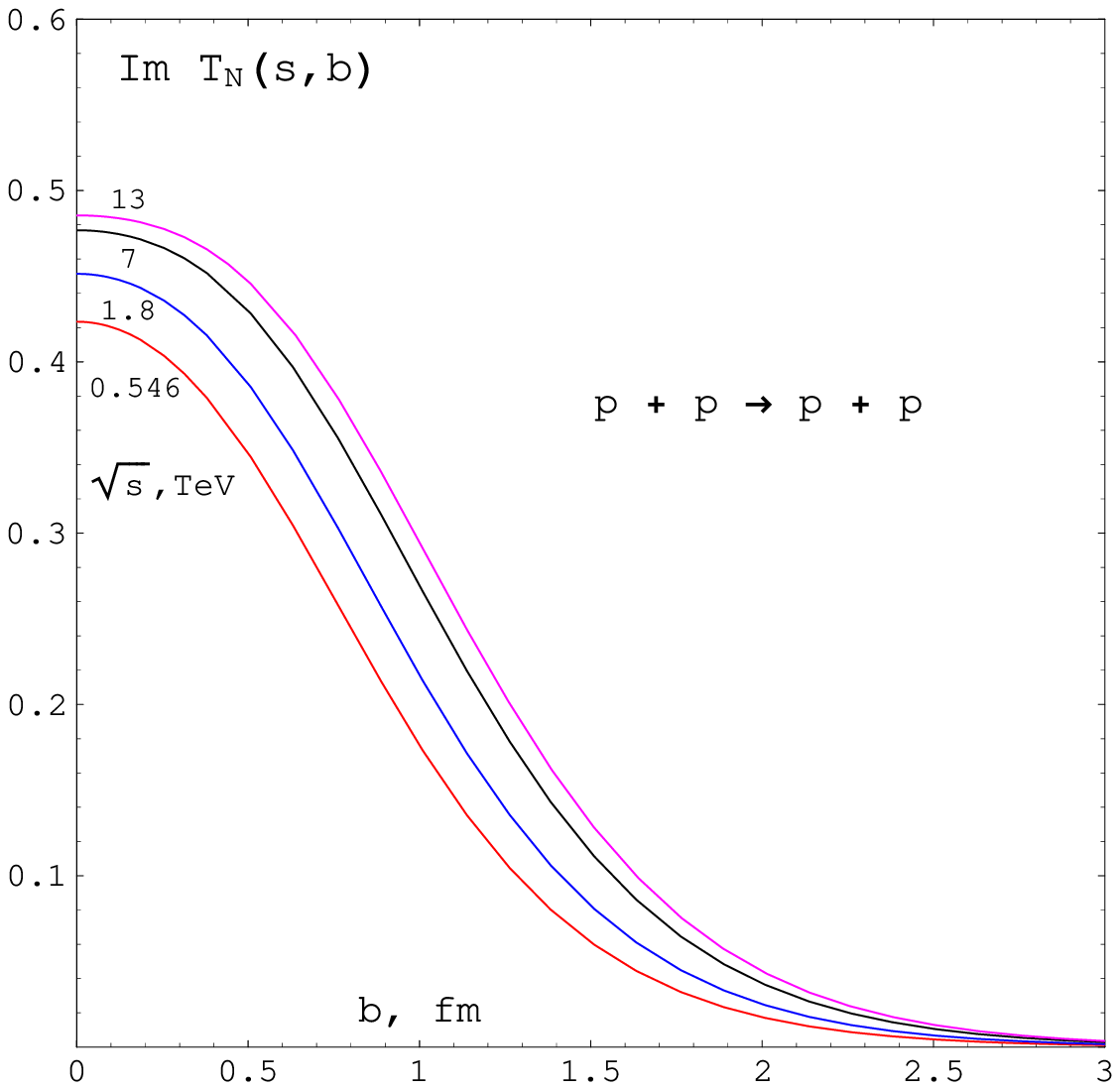}
\vskip -8.3cm
\hskip 8.65cm
\epsfxsize=8.35cm\epsfysize=8.35cm\epsffile{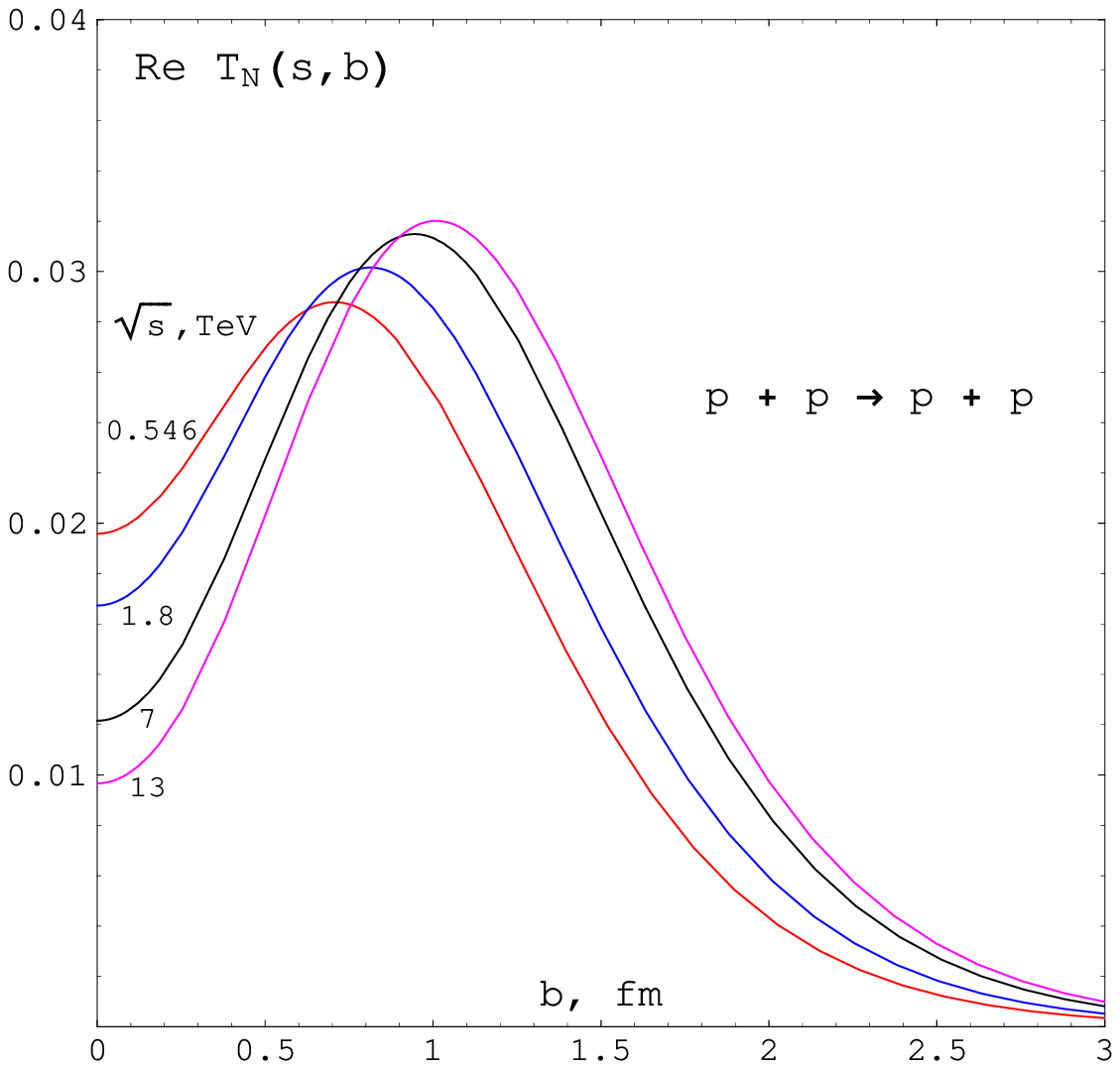}
\caption{The imaginary and real parts of the strong interaction subamplitude $T_N(s,b)$ at various values of the collision energy.}
\label{coord}
\end{figure}

At large enough values of the impact parameter, electromagnetic interaction dominates: $|\delta_C(s,b)|\gg |T_N(s,b)|$ and, hence, $T(s,b)\approx \delta_C(s,b)$. In its turn, 
at small values of $b$ the electromagnetic interaction of protons can be ignored: $T(s,b)\approx T_N(s,b)$. In the range\linebreak 3 fm $<b<$ 10 fm, wherein 
$|\delta_C(s,b)|\sim |T_N(s,b)|$, we may ignore the third term in (\ref{elnuc}), because $|T_N(s,b)|\ll 1$ in this region (the fast decrease of $|T_N(s,b)|$ can be seen in 
Fig. \ref{coord}), and, consequently, this term is negligible as compared with the first two.

Thus, finally, we come to the leading approximation of the full (electromagnetic + strong) amplitude in the entire kinematic range of EDS:
\begin{equation}
\label{elnuc2}
T(s,b)\approx\delta_C(s,b)+T_N(s,b)\,,\;\;\Rightarrow \;\;T(s,t)\approx\delta_C(s,t)+T_N(s,t)\,.
\end{equation}
In other words, we neglect the Coulomb-nuclear interference in the amplitude level.

It should be noted here that such a negligibility is a model-dependent effect. In the framework of many models, the corresponding terms are expected to yield a significant 
contribution into the full amplitude (for detailed discussion, see, say, \cite{petrov} and references therein).

In the transferred momentum range $0<\sqrt{-t}<2$ GeV, the Coulomb term can be approximated by a simple expression 
\begin{equation}
\label{coulomb}
\delta_C(s,t) = \pm\frac{8\,\pi\,s\,\alpha_e}{t}F_E^2(t)\,,
\end{equation}
where $F_E(t)=\left(1-\frac{t}{0.71\,\rm GeV^2}\right)^{-2}$ is the dipole electric form-factor of proton.

\section*{3. Verification of the model}

\subsection*{3.1. The model predictions {\it versus} the newest experimental data}

\begin{figure}[ht]
\vskip -0.3cm
\epsfxsize=8.2cm\epsfysize=8.2cm\epsffile{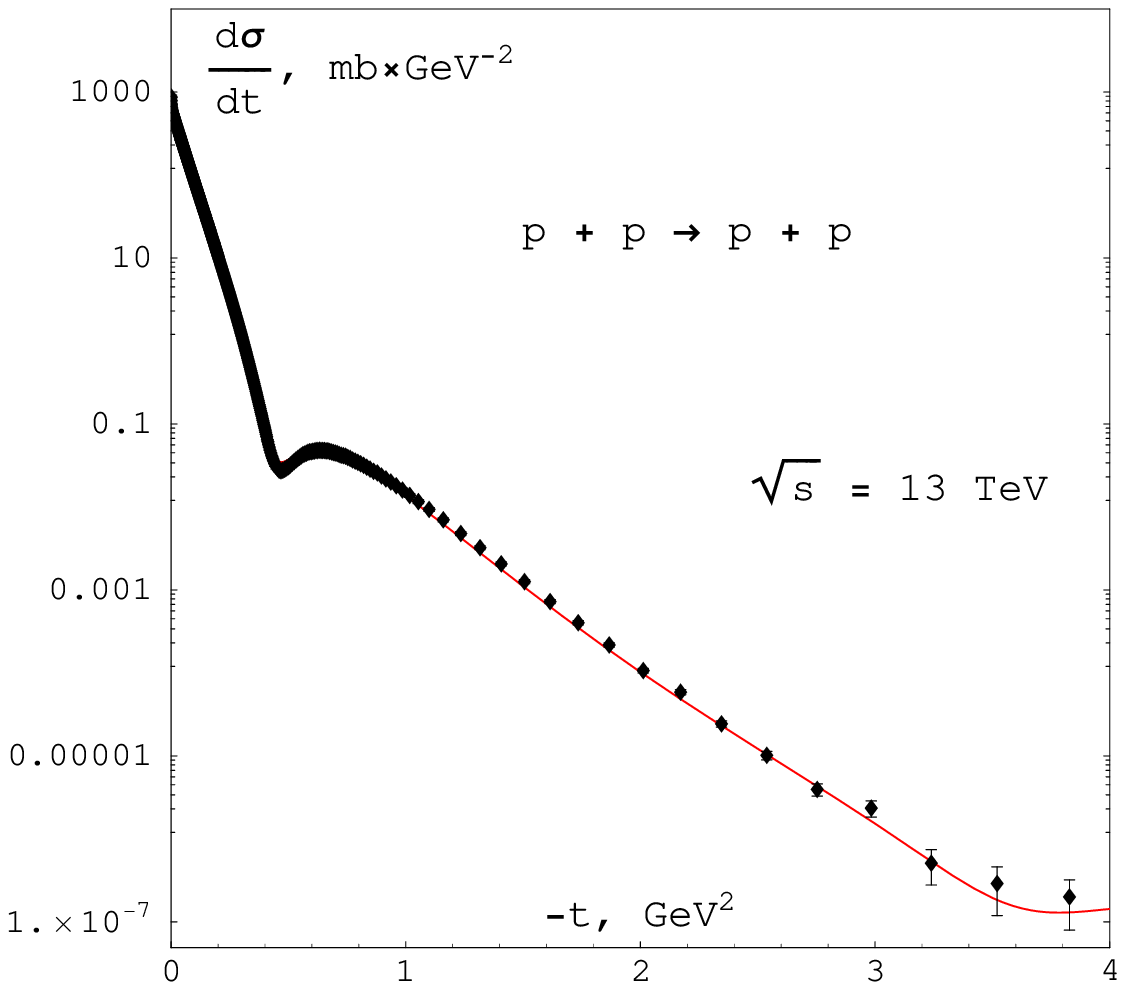}
\vskip -8.1cm
\hskip 8.9cm
\epsfxsize=7.9cm\epsfysize=7.9cm\epsffile{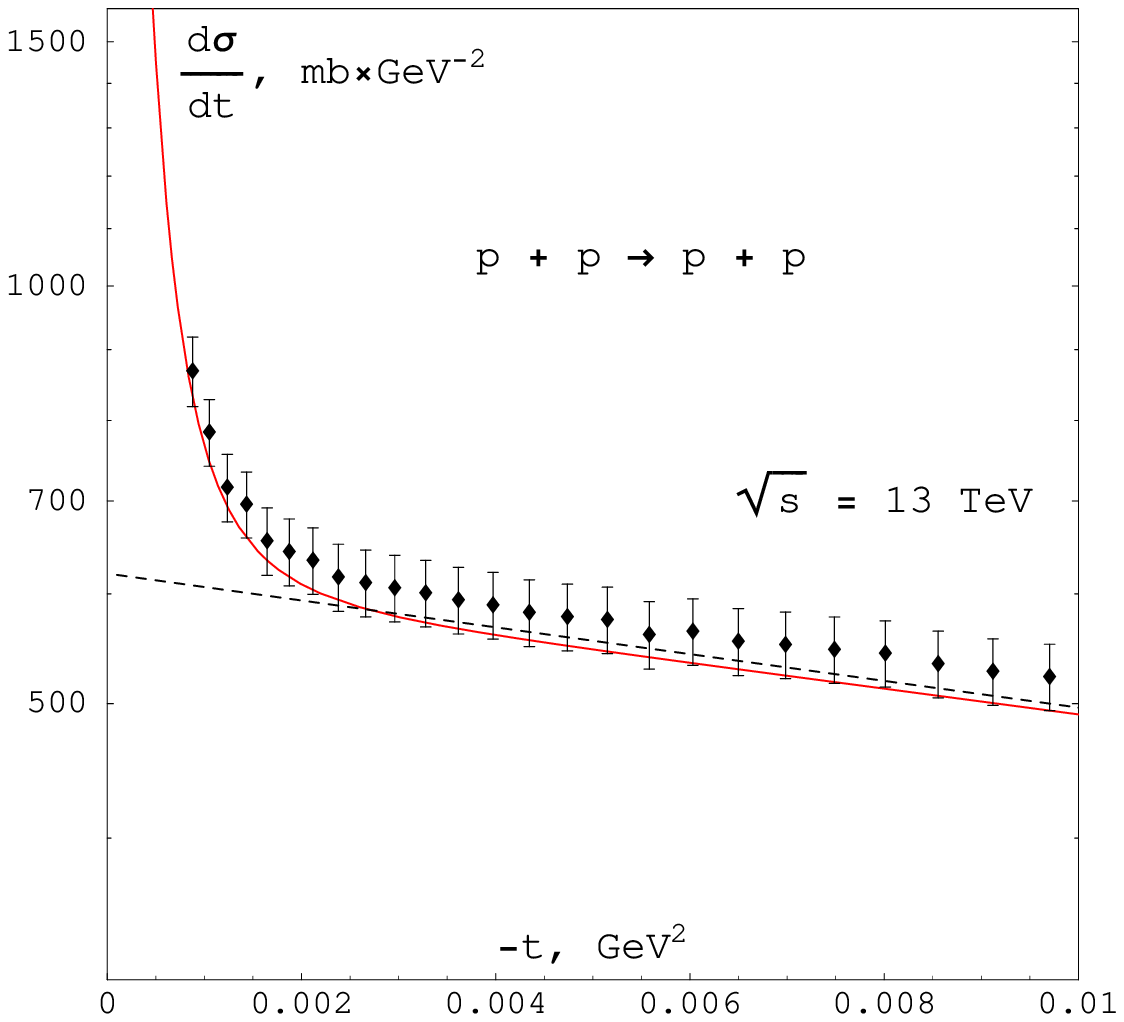}
\vskip -0.3cm
\caption{The predictions of the model \cite{godizov} in the case $\alpha_{\rm HP}(0)-1=0.32$ {\it versus} the TOTEM data at $\sqrt{s}=$ 13 TeV \cite{totem13}. The dashed line 
corresponds to the approximation $\delta_C(s,t)=0$.}
\label{pred}
\end{figure}

To check the model efficiency, we need to compare the model predictions with the new data \cite{totem13} on the proton-proton EDS at $\sqrt{s}=$ 13 TeV. The results of 
such a verification (without any refitting of the model parameters) are presented in Fig. \ref{pred}. 

The data description quality in terms of the method of least squares is $\chi^2=$ 1796 over 428 points (the description quality in the range 
$\sqrt{-t}<$ 0.1 GeV is $\chi^2=$ 18 over 25 points). Hereby, we observe a rather weak deviation of the model curve from the experimental data.\footnote{It should be pointed 
out here that the exploited parametrizations (\ref{pomeron}) for $\alpha_{\rm SP}(t)$ and $g_{\rm SP}(t)$ allow one to obtain a satisfactory description of the exclusive 
photoproduction of light vector mesons \cite{godizov5} and the proton single diffractive dissociation \cite{godizov3} in the kinematic ranges where the impact of 
the hard Pomeron and secondary Reggeon exchanges is negligible as compared with the experimental uncertainties.}

The model predictions for the $pp$ total cross-section and for $\rho = \frac{{\rm Re}\,T_N(s,0)}{{\rm Im}\,T_N(s,0)}$ at $\sqrt{s}=$ 13 TeV are 
$\sigma^{model}_{tot}$(13 TeV) $\approx$ 109.4 mb and $\rho^{model}$(13 TeV) $\approx$ 0.125, while the corresponding measured values are 
$\sigma_{tot}$(13 TeV) = (110.5\,$\pm\,$2.4) mb and $\rho\,$(13 TeV) = $0.10\pm0.01$ \cite{totem13}. It should be noted here that extraction of these quantities from the 
experimental angular distributions is a strongly model-dependent procedure.
\begin{figure}[ht]
\begin{center}
\vskip -0.8cm
\epsfxsize=8cm\epsfysize=8cm\epsffile{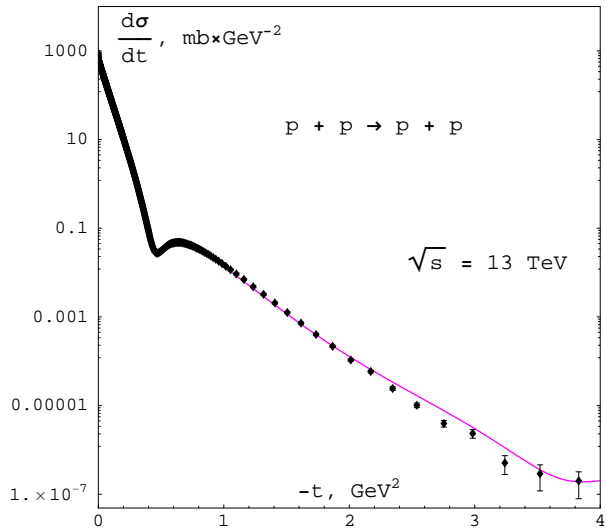}
\end{center}
\vskip -0.7cm
\caption{The prediction of the model \cite{godizov} in the case $\alpha_{\rm HP}(0)-1=0.44$ {\it versus} the TOTEM data at $\sqrt{s}=$ 13 TeV \cite{totem13}.}
\label{pred2}
\end{figure}

Concerning the sensitivity of the considered model with respect to the fixed value of the HP intercept, it should be noted here that if we put $\alpha_{\rm HP}(0)=1.44$ 
and fit $\beta_{\rm HP}(t)$ to the data at 7 TeV of the collision energy (see Fig. 2 in \cite{godizov}), then, at 13 TeV, we obtain the prediction for the $pp$ angular 
distribution presented in Fig. \ref{pred2}. As we can see, at such a value of the HP intercept, the model demonstrates a noticeable overestimation of the observed 
$d\sigma/dt$ in the region\linebreak $\sqrt{-t}>$ 1.5 GeV. In other words, the proposed phenomenological scheme is quite sensitive to the value of $\alpha_{\rm HP}(0)$ and, 
thus, we may interpret the result presented in Fig. \ref{pred} as some model-dependent confirmation of the fact that the HP in the high-energy EDS of nucleons is the same 
supercritical Reggeon that governs the low-$x$ behavior of the proton unpolarized structure function $F^p_2(x,Q^2)$ at high values of the incoming photon virtuality $Q^2$
in deeply inelastic scattering of leptons on protons. 

\subsection*{3.2. The results of refitting}

\begin{figure}[ht]
\epsfxsize=8.2cm\epsfysize=8.2cm\epsffile{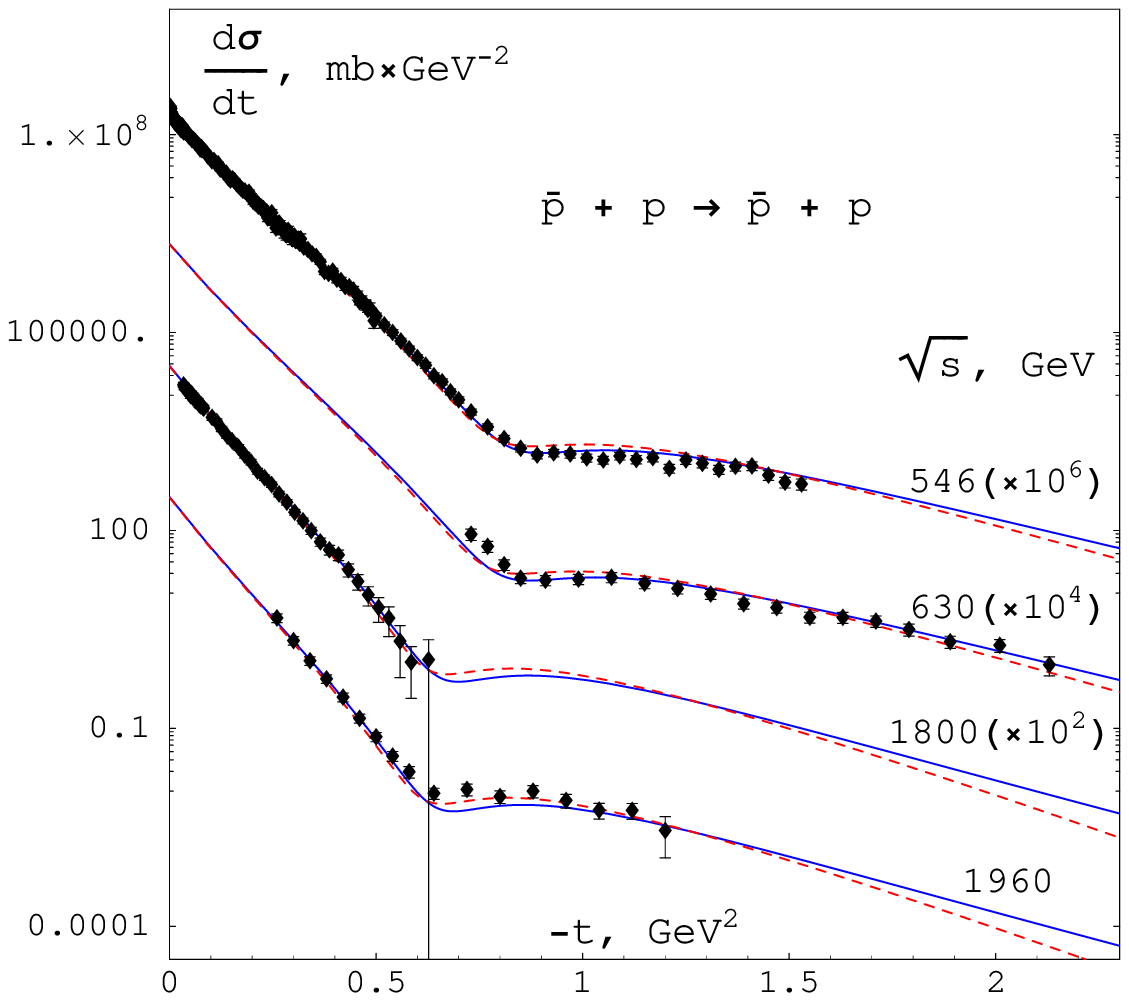}
\vskip -8.25cm
\hskip 8.7cm
\epsfxsize=8.25cm\epsfysize=8.25cm\epsffile{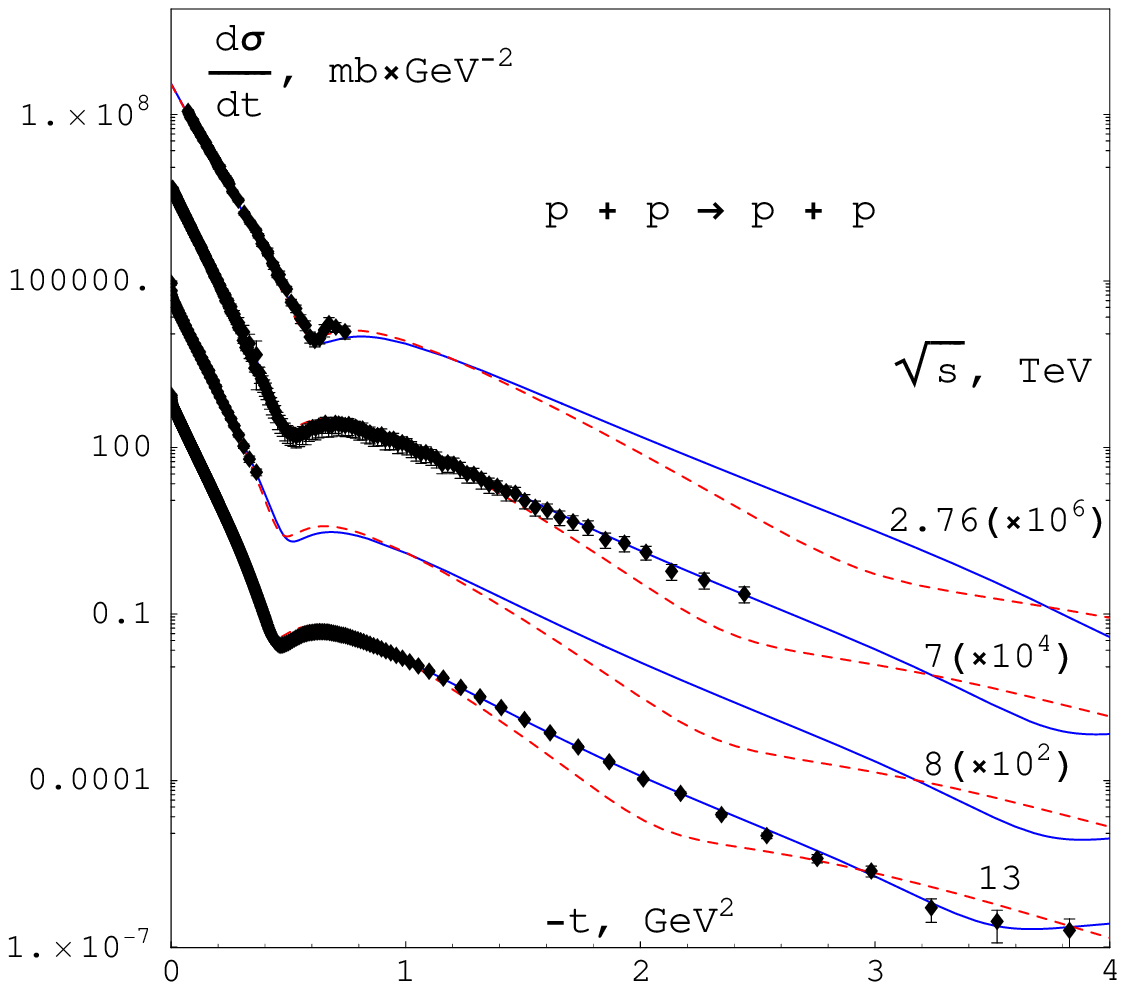}
\vskip -0.3cm
\caption{The differential cross-sections of nucleon-nucleon EDS at ultrahigh energies. The dashed lines correspond to the one-Pomeron eikonal approximation wherein the HP 
exchange contribution to the eikonal is ignored.}
\label{refit}
\end{figure}

The next step is to refit the model parameters to the enlarged set of data, exploiting the same parametrizations (\ref{pomeron}) for the unknown functions 
$\alpha_{\rm SP}(t)$, $g_{\rm SP}(t)$, and $\beta_{\rm HP}(t)$ as in \cite{godizov} (it is necessary for improvement of the description quality). The results are presented 
in Tables \ref{tab2} and \ref{tab3} and Fig. \ref{refit}. 

The main cause of the observed slight discrepancy between the model issues and the experimental data is, certainly, the stiffness of the used parametrizations for the SP 
Regge trajectory and the SP coupling to proton. More flexible parametrizations could result in a better description of the data, though one should keep in mind that the true 
analytic behavior of $\alpha_{\rm SP}(t)$, $g_{\rm SP}(t)$, and $\beta_{\rm HP}(t)$ remains unknown. Nonetheless, the simplicity of test functions (\ref{pomeron}) makes them 
very attractive for usage in the region of low negative $t$, while the achieved quality of the description makes the model quite suitable for rough estimations and predictions 
of the nucleon-nucleon EDS observables at ultrahigh energies.

\begin{table}[ht]
\begin{center}
\begin{tabular}{|l|l|}
\hline
\bf Parameter          & \bf Value         \\
\hline
$\alpha_{\rm SP}(0)-1$  & 0.1055           \\
$\tau_a$                & 0.572 GeV$^2$    \\
$g_{\rm SP}(0)$         & 14.7 GeV         \\
$a_g$                   & 0.32 GeV$^{-2}$  \\
$\beta_{\rm HP}(0)$     & 0.108            \\
$b$                     & 1.55  GeV$^{-2}$ \\
$\alpha_{\rm HP}(0)-1$  & 0.32 (FIXED)     \\
\hline
\end{tabular}
\end{center}
\caption{The parameter values for (\ref{pomeron}) obtained via fitting to the EDS data in the entire kinematic range \{546 GeV $\le\sqrt{s}\le$ 13 TeV, $\sqrt{-t}<$ 2 GeV\}.}
\label{tab2}
\end{table}

\begin{table}[ht]
\begin{center}
\begin{tabular}{|l|l|l|}
\hline
$\sqrt{s}$, GeV                         & \bf Number of points &  $\chi^2$   \\
\hline
  546 ($\bar p\,p$; UA4)                &  187                 &  264        \\
  630 ($\bar p\,p$; UA4)                &   19                 &   34        \\
 1800 ($\bar p\,p$; E710)               &   51                 &   19        \\
 1960 ($\bar p\,p$; D0)                 &   17                 &   24        \\
 2760 ($p\,p$; TOTEM)                   &   63                 &  180        \\
 7000 ($p\,p$; TOTEM, ATLAS)            &  205                 &  323        \\
 8000 ($p\,p$; TOTEM, ATLAS)            &   69                 &  159        \\
13000 ($p\,p$; TOTEM)                   &  428                 &  982        \\
\hline
\bf Total                               & 1039                 & 1985        \\
\hline
\end{tabular}
\end{center}
\vskip -0.2cm
\caption{The quality of the description of the data \cite{totem7,totem13,elastic} on the EDS angular distributions.}
\label{tab3}
\end{table}

Another possible cause of the noticeable deviation of the model curves from the $\bar pp$ scattering data at $\sqrt{s}=$ 1.96 TeV and 0.6 GeV$^2<-t<$ 0.9 GeV$^2$ and, in 
the same $t$-interval, from the $pp$ scattering data at $\sqrt{s}=$ 2.76 TeV is the impact of exchanges by the Odderon (the $C$-odd counterpartner of the SP). The Odderon 
exchange contribution is completely ignored in the framework of the considered two-Pomeron eikonal approximation. Although almost nobody doubts the existence of such a 
supercritical Reggeon which gives different contributions into the $pp$ and $\bar pp$ interactions, the significance of the Odderon exchanges at available energies is still 
a subject of discussion. For example, in the most known model of elastic diffractive scattering \cite{landshoff} the influence of the Odderon exchanges is considered to be 
negligible. On the contrary, the authors of paper \cite{martynov} insist on the necessity to take account of the Odderon contribution. The scheme presented in \cite{khoze} 
admits both the variants. In any case, the Odderon impact on the angular distributions of the high-energy elastic scattering of nucleons seems to be much more fine effect 
than the impact of the HP exchanges. At the LHC energies, ignoring of the HP contribution into the eikonal can lead to catastrophic divergence between the model curves and 
the data in the region $-t>$ 1.5 GeV$^2$ (see Fig. \ref{refit}).

\section*{4. Conclusion}

In the light of the aforesaid, we may conclude that the two-Pomeron eikonal approximation successfully confirmed its predictive value and, thus, it can be considered as a 
simple and reliable phenomenological tool for qualitative description of the nucleon-nucleon EDS at ultrahigh energies. As well, it was confirmed in the framework of the 
proposed model that the HP which has a crucial impact on the nucleon-nucleon diffraction pattern at the LHC energies and transferred momenta higher than 1 GeV is the same 
supercritical Reggeon which governs the  low-$x$ evolution of the proton unpolarized structure function $F^p_2(x,Q^2)$ in DIS at high values of the incoming photon virtuality. 
The test functions $\alpha_{\rm SP}(t)$ and $g_{\rm SP}(t)$ with the free parameter values fitted to the available data can be used in the framework of Reggeon models for 
more compound reactions, such as high-energy single diffractive dissociation (SDD) of nucleons or central exclusive production (CEP) of light neutral mesons.

\end{document}